\documentstyle[12pt]{article}
\begin{document}
%
\def    \be             {\begin{eqnarray}}
\def    \ee             {\end{eqnarray}}
%

\begin{flushright}
Freiburg--THEP 96/4\\
February 1996
\end{flushright}

\vspace{1.5cm}

\begin {center}
{\large \bf  The hidden Higgs model at the NLC
\footnote{Contribution to the LC2000 workshop of the European
working group on Physics with $e^+e^-$ linear colliders.}}

\vskip 1.cm
{\bf T. Binoth},  
{\bf J. J. van der Bij} \\ 
\bigskip
Albert--Ludwigs--Universit\"at Freiburg, 
Fakult\"at f\"ur Physik, \\
Hermann--Herder--Strasse 3, 79104 Freiburg i. Br.,
\end{center}

\medskip

\begin{abstract}
We investigate the influence of massless scalar singlets on Higgs signals
at the NLC. An exclusion bound is presented which restricts large regions of 
the parameter space but on the other hand implies that for strong 
interactions between the Higgs boson and the singlet 
fields of the hidden sector, detection of such a non standard Higgs 
signal can become impossible. 
\end{abstract}

\medskip

\section{Introduction}

Understanding of the electroweak symmetry breaking mechanism is one of the main
tasks in particle physics. The determination  of its nature would be 
a break-through in our knowledge about matter. So it is important to 
think about alternatives to the Standard Model Higgs sector. Various such 
extensions are available. Maybe the best motivated one is the supersymmetrized 
Standard Model with its important phenomenological implication of a light Higgs 
boson and which allows a consistent frame for grand unified theories. Another 
well understood extension -- though in its minimal version disfavoured by the 
precision experiments at LEP -- are
technicolor theories. Though these theories avoid fundamental scalars, a rich
bosonic spectrum of techniquark condensates may exist. Thus in both theories, 
as long as they do not occur in their minimal form, light bosonic matter could 
be present modifying the standard Higgs signals we are looking for at present 
and future colliders. If such bosons appear as singlets under the Standard 
Model gauge group, they do not feel the color or electroweak forces,
but they can couple to the Higgs particle. As a consequence
radiative corrections to weak processes are not sensitive to the
presence of singlets in the theory, because no Feynman graphs containing
singlets  appear
at the one--loop level. Since effects at the two--loop level
are below the experimental precision,
the presence of a singlet sector is not ruled out by any 
of the LEP1 precision data. The only connection to such a hidden sector
is a possible Higgs--singlet coupling, leading to a nonstandard invisible
Higgs decay.
The invisible decay of the Higgs boson with a narrow width 
leads to relatively sharp missing energy signals, well known from discussions
on Majoron models \cite{valle}. However a strongly coupled hidden sector 
could lead to fast Higgs decay and thereby to wide resonances. 
This would disturb the signal to background 
ratio if necessary cuts are imposed.  

To check the influence of a hidden sector we will study the coupling
of a Higgs boson to an O(N) symmetric set of scalars, which  
is one of  the simplest possibilities, introducing only a few extra 
parameters in the theory. The effect of the extra scalars is practically
the presence of a possibly large invisible decay width of the Higgs particle.
When the coupling is large enough the Higgs resonance can become
wide even for a light Higgs boson. It was shown earlier that there
will be a range of parameters, where such a Higgs boson can be seen neither
at LEP nor at the LHC \cite{vladimir,lep2report}. 

In the next section we will introduce the 
model together with its theoretical constraints and in the last section
we will discuss exclusion limits at the NLC.  

\section{The model}
The scalar sector of the model consists of the usual Higgs sector coupled 
to a real N--component vector $\vec\varphi$ of scalar fields, denoted by 
phions in the following. The Lagrangian density is given by,
\be
\label{definition}
 {\cal L}  &=&
 - \partial_{\mu}\phi^+ \partial^{\mu}\phi -\lambda (\phi^+\phi - v^2/2)^2
   - 1/2\,\partial_{\mu} \vec\varphi \partial^{\mu}\vec\varphi
     -1/2 \, m^2 \,\vec\varphi^2 \nonumber \\
     &&- \kappa/(8N) \, (\vec\varphi^2 )^2
    -\omega/(2\sqrt{N})\, \, \vec\varphi^2 \,\phi^+\phi \nonumber
\ee
where $\phi$ is the standard Higgs doublet. 
Couplings to fermions and vector bosons are the same as in the Standard Model.
The ordinary
Higgs field acquires the vacuum expectation value $v/\sqrt{2}$. For positive 
$\omega$ the $\vec\varphi$--field acquires no vacuum expectation
value. After spontaneous
symmetry breaking one is left with the ordinary Higgs boson,
coupled to the phions into which it decays. Also the phions
receive an induced mass from the spontaneous symmetry breaking which is 
suppressed by a factor $1/\sqrt{N}$.
If the factor N is taken
to be large,  the model can be analysed with $1/N$--expansion techniques.
By taking this limit the phion mass remains small, but as there
are many phions, the decay width of the Higgs boson can become large.
Therefore the main effect of the presence of the phions is to give
a large invisible decay rate to the Higgs boson. The 
invisible decay width is given by 
\be \Gamma_H =\frac {\omega^2 v^2}{32 \pi M_H} = 
\frac{\omega^2 (\sin\theta_W\cos\theta_W M_Z)^2}{32 \pi^2 \alpha_{em} M_H}
\quad .\nonumber \ee
The Higgs width is compared with the width in the Standard Model for various 
choices of the coupling $\omega$ in Fig.~\ref{width}.
The model is different
from Majoron models \cite{valle}, since the width is not necessarily small.
The model is similar to the technicolor--like model of Ref.~\cite{chivukula}.
\begin{figure}[hbt]
\vspace{0.1cm}
\caption{\it Higgs width in comparison with the Standard Model.}
\label{width}
\end{figure}

Consistency of the model requires two conditions.
One condition is the absence of a Landau pole below a certain scale
$\Lambda$. The other follows from the stability of the vacuum up to a certain
scale. An example of such limits is given in Fig.~\ref{stability},
where $\kappa=0$ was taken at the scale $2m_Z$, which allows for
the widest parameter range.
The regions of validity up to a given scale $\Lambda$ are sandwiched
between the lower--left and the upper--right contour lines in the figure. 
The first stem from instability of the 
vacuum, the second from the presence of a Landau pole at that scale.

\begin{figure}[htb]
\vspace{0.1cm}
\caption{\it Theoretical limits on the parameters of the model
in the $\omega$ vs. $M_H$ plane. The contour lines correspond 
to the cutoff scales $\Lambda = 10^{19}$, $10^6$, $10^4$ and $10^3$ GeV.}
\label{stability}
\end{figure}

To search for the Higgs boson there are basically
two channels, one is the standard decay, which is reduced in branching
ratio due to the decay into phions.
The other is the invisible decay, which rapidly becomes dominant,
eventually making the Higgs resonance wide (see Fig.~\ref{width}). 
In order to give the bounds we 
neglect the coupling $\kappa$ as this is a small effect. We
also neglect the phion mass. For other values of the phion mass
the bounds can be found by rescaling the decay widths
with the appropriate phase space factor. Now we confront this
two dimensional parameter space with the experimental potential 
of the NLC.  

\section{NLC bounds}
At the NLC the upper limits on the couplings in the present model
come essentially from the invisible decay, as the branching ratio
into visible particles drops with increasing $\varphi$--Higgs
coupling ($\omega$), whereas for small $\omega$ one has to consider visible 
Higgs decays, too. Since the main source for Higgs production, the 
$WW$--fusion process, can not be used to look for invisible Higgs decay,
one is in principle left with the Higgsstrahlung und $ZZ$--fusion reaction. 
For energies up to 500 GeV the Higgsstrahlungs cross section is dominant and
is of comparable size to the $ZZ$--fusion process even if one is folding in 
the branching ratio $B(Z\rightarrow e^+e^-,\mu^+\mu^-)$. 
The possibility to tag an on--shell Z boson via a leptonic system which is 
extremely useful for the discrimination of possible backgrounds makes 
Higgsstrahlung to be the preferred production mechanism. Thus we only have 
considered reactions containing an on shell Z boson with its decay into 
$e^+e^-$ or $\mu^+\mu^-$. One should be aware that a few events from the 
huge $WW$ background may survive \cite{eboli}, but that the $Z\nu\nu$ 
background is dominant after imposing the cuts defined below.
Then the signal cross section is the well  known Higgsstrahlungs cross section 
modified by the non standard Higgs width due to phion decay. With the invariant 
mass of the invisible phion system, $s_I$, it has the form:  
\be \sigma_{(e^+e^-\rightarrow Z+E\!\!\!/)} = 
\int ds_I \, \sigma_{(e^+e^-\rightarrow ZH)}(s_I) \,
\frac{\sqrt{s_I} \quad \Gamma(H\rightarrow E\!\!\!/)}
{\pi ((M_H^2-s_I)^2+s_I\,\Gamma(H\rightarrow \mbox{All})^2)}\nonumber\ee
We calculated the $Z\nu\nu$ background with the standard set of graphs for
Z production ($ZZ$--production, $WW$--fusion and Z initial, final state 
radiation) by a Monte Carlo program (see Ref.~\cite{mele}). To reduce the 
background we used the fact that the angular distribution of the Z--boson 
for the signal peaks for small values of $|\cos\theta_Z|$ in contrast to the 
background. Thus we imposed the cut $|\cos\theta_Z|<0.7$.
Because we assume the reconstruction of the on-shell Z--boson we use the 
kinematical relation $E_Z=(s+M_Z^2-s_I)/(2\sqrt{s})$
between the Z energy and the invariant mass of the invisible system
to define a second cut. Since the differential cross section $d\sigma/ds_I$ 
contains the Higgs resonance at $s_I=M_H^2$, we impose the following condition 
on the Z energy:
\be \frac{s+M_Z^2-(M_H+\Delta_H)^2}{2\sqrt{s}}<E_Z<
\frac{s+M_Z^2-(M_H-\Delta_H)^2}{2\sqrt{s}} \nonumber\ee 
For the choice of $\Delta_H$ a comment is in order. As long as the Higgs width 
is small, one
is allowed to use small  $\Delta_H$, which reduces the background considerably 
keeping most of the signal events. But in the case of large $\varphi$--Higgs 
coupling, $\omega$, one looses valuable events. To compromise between both 
effects we took  $\Delta_H=30$ GeV.  
 
For the exclusion limits we assumed an integrated luminosity
of $20\,fb^{-1}$. To define the $95 \%$ confidence level we used 
Poisson statistics similar to the description of Ref. \cite{lep2report}.
The result is given  
in Fig.~\ref{exclusion}. One notices the somewhat reduced sensitivity
for $M_H\simeq M_Z$ due to a resonating $Z$ boson in the $ZZ$ background. For 
larger values of $M_H$ the limit stems from the other $Z\nu\nu$ backgrounds
with $W$ bosons in the t--channel and kinematical constrains. 
For large $\omega$ the signal ceases
to dominate over the background because the Higgs peak is smeared out
to an almost flat distribution. 

\begin{figure}[htb]
\vspace{0.1cm}
\caption{\it Exclusion limits at the NLC due to Higgs searches. The dashed
line corresponds to the invisible, the full line to all Higgs decay modes.}
\label{exclusion}
\end{figure}

We conclude from this analysis that the NLC can put further
limits on the parameter space of our invisible Higgs model. 
Note that within the kinematic range very strong limits on $\omega$ can be set. 
Again there is a range
where the Higgs boson will not be discovered, even if it does exist in this 
mass range. This has already been shown for the Higgs search at LEP and also
holds true for the heavy Higgs search at LHC. 
We see that a sufficiently wide nonstandard Higgs resonance would make it very 
difficult to test the mechanism of electroweak symmetry breaking at 
future colliders. 

\end{document}